\documentclass{article}
\usepackage{amsmath,amssymb,amsthm,bm,helvet}
\usepackage{graphicx}

\newcommand{\ve}[1]{\bm{#1}} 
\newcommand\tit[1]{``#1,''}
\newcommand{\jl}{\em}

\newcommand{\ie}{{\it i.e.},}
\newcommand{\eg}{{\it e.g.},}
\newcommand{\Eq}[1]{Eq.~(\ref{#1})}
\newcommand{\Sec}[1]{Sec.~\ref{Sec:#1}}
\newcommand{\Fig}[1]{Fig.~\ref{Fig:#1}}
\newcommand{\App}[1]{Appendix \ref{App:#1}}

\newcommand{\FI}[1]{\mathcal{I}_{#1}}
\newcommand{\FII}[1]{\mathcal{I}_{#1}^{(\mathsf I)}}
\newcommand{\FIII}[1]{\mathcal{I}_{#1}^{(\mathsf{II})}}
\newcommand{\DI}[1]{\mathcal{D}_{#1}}

\newcommand{\cld}{\mu}
\newcommand{\seg}{\iota}
\newcommand{\dst}{\eta}
\newcommand{\cor}{\gamma}
\newcommand{\Cld}{F_\cld}
\newcommand{\Seg}{F_\seg}
\newcommand{\Dst}{F_\dst}
\newcommand{\av}[1]{\langle{#1}\rangle}
\newcommand{\CB}{\mathfrak{V}} 
\newcommand{\NB}{\mathfrak{N}} 
\newcommand{\Prob}{\mathsf{P}}
\newcommand{\R}{\mathbb{R}}
\newcommand{\Sph}{\mathbb{S}}
\newcommand{\T}{\mathcal{T}}
\newcommand{\opt}[2]{\mathcal{O}_{#1}^{#2}\!}
\newcommand{\FJ}{\mathcal{J}}
\newcommand{\G}{G}
\newcommand{\Rd}{\mathsf R}
\newcommand{\Lh}{\mathsf L}
\newcommand{\Wd}{\mathsf W}
\newcommand{\Fd}{\mathsf F}
\newcommand{\TRI}{\blacktriangle}
\newcommand{\SQU}{\blacksquare}
\newcommand{\LLint}[1]{(\Lh\cap{#1})_\rightthreetimes^2}
\newcommand{\nI}{n_{\rm I}}

\newcommand{\sigseg}{\seg_\pm}
\newcommand{\sigcld}{\cld_\pm}
 
\newtheorem{Lemm}{Lemma}
\newtheorem{Defn}{Definition}

\newcommand{\lemm}[1]{{\it Lemma~\ref{#1}}}
\newcommand{\defn}[1]{{\it Definition~\ref{#1}}}
\numberwithin{equation}{section}

\newcommand{\scrf}{\fontfamily{cmfr}\selectfont}

\title{\sf\bfseries Signed Chord Length Distribution\\
 Part I}
\date{}
\author{\sf\itshape Alexander \t{Yu}.~Vlasov}
\begin{document}
\sloppy

\maketitle

\makeatletter
\global\@specialpagefalse
\renewcommand{\@oddfoot}{\rlap{\copyright{\it\scrf
\ A.\t{Yu}.Vlasov,} 2007} \hfil \thepage \hfil \llap{\sf Signed CLD}}
\renewcommand\section{\@startsection {section}{1}{\z@}%
                                   {-3.5ex \@plus -1ex \@minus -.2ex}%
                                   {2.3ex \@plus.2ex}%
                                   {\normalfont\large\sf\bfseries\boldmath\uppercase}}
\renewcommand\subsection{\@startsection{subsection}{2}{\z@}%
                                     {-3.25ex\@plus -1ex \@minus -.2ex}%
                                     {1.5ex \@plus .2ex}%
                                     {\normalfont\sf\bfseries\itshape\boldmath}}
\let\@ldappendix=\appendix
\renewcommand\appendix{\@ldappendix%
  \def\pref@section{\uppercase{\appendixname}~}%
  \def\@seccntformat##1{\csname pref@##1\endcsname\csname the##1\endcsname\quad}%
  \renewcommand{\theequation}{\thesection\arabic{equation}}%
  \renewcommand{\thesubsection}{\thesection-\arabic{subsection}}%
}

\makeatother

\begin{abstract}
 In this paper is discussed an application of signed measures (charges) 
to description of segment and chord length distributions in nonconvex bodies. 
The signed distribution may naturally appears due to definition via derivatives
of nonnegative autocorrelation function simply related with distances distribution 
between pairs of points in the body. In the work is suggested constructive
geometrical interpretation of such derivatives and illustrated appearance 
of ``positive'' and ``negative'' elements similar with usual Hanh--Jordan
decomposition in measure theory. The construction is also close related with
applications of Dirac method of chords. 
\end{abstract}

{\sf\boldmath\tableofcontents}

\section{Introduction}
\label{Sec:Intro}

\begin{figure}[hbt]
\begin{center}
\includegraphics[scale=0.5]{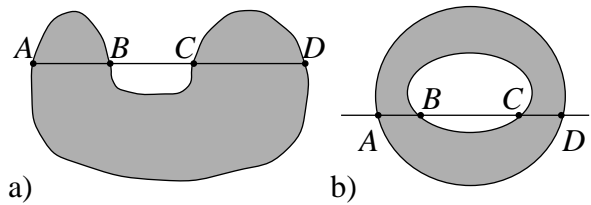}
\end{center}
\caption{Nonconvex bodies: a) simply connected b) with hole.}\label{Fig:nonconv}
\end{figure}

Different properties of chord length distribution (CLD) for nonconvex bodies were 
discussed recently in few  publications \cite{Gil00,MRG03,MRD03,BR01,Str01,Han03,GMR05}.
Let us recall three different ways to introduce CLD for nonconvex body.
Straight line may intersect nonconvex body more than one time (see \Fig{nonconv})
and we can either consider each segment of such line as separate chord or calculate 
sum of all such segments. These two methods are known as multi-chord and one-chord 
distribution (MCD and OCD) respectively \cite{Gil00,MRG03,MRD03}.

For the convex case probability density function for CLD is proportional to second 
derivative of autocorrelation function \cite{Gil00,BR01,Str01,Han03}. 
Such a property may be used for a third definition of ``the generalized  
chord distribution'' \cite{BR01,Str01,Han03}, but straightforward 
calculations demonstrates possibility of negativity of such function 
for some nonconvex bodies \cite{Han03}. Let us call this function here
{\em signed chord (length) distribution} to avoid some ambiguity of term
``generalized'' and to emphasize the basic distinguishing property 
of this function.

Definition of CLD for convex body has standard {\em probabilistic interpretation}
in theory of {\em geometric probability} and {\em random sets} \cite{Ken,San,Mat}.
The MCD and OCD cases for nonconvex body may be described as well \cite{Gil00,MRG03,MRD03}. 
Is it possible to consider similar possibility for {\em the signed chord distribution}?

In wonderful essay {\it ``Negative probability''} \cite{Fey87} Feynman wrote
that, unlike {``final probability of a verifiable physical event''}, 
{``conditional probabilities and probabilities of imagined intermediate states 
may be negative''} and so: {\em ``If a physical theory for calculating probabilities 
yields a negative probability for a given situation under certain assumed conditions, 
we need not conclude the theory is incorrect.''}
In this review Feynman provided a few examples with appearance and interpretation 
of negative probabilities both for quantum and classical physical models. 

Mathematical extension of the measure
theory for such a purposes may use so-called {\em signed measures (charges)} \cite{KolmFom}.
 Usually such extension is reduced to standard positive measures 
due to {\em Hahn and Jordan decompositions}, corresponding to expression of 
charge as difference of two positive measures \cite{KolmFom}.

In many processes with signed distributions the Hahn decomposition, {\ie}
splitting of space of events on positive and negative parts is
quite obvious, {\eg} in simplest examples we have two kinds of events: 
putting and removing objects \cite{Fey87}.
A distinction of {\em the signed chord distribution} is appearance of negativity 
due to differentiations of positive function without such a natural decomposition 
on positive and negative elements.

Nonconvex body \Fig{nonconv}b may be represented as a convex body 
and a convex hole and it provides some intuitive justification of
possibility to express some distributions using formal difference of 
convex hull and the hole. Rigor consideration is more difficult, 
especially for chord distribution expressed via second derivative, {\eg}
method derived below in \Sec{SigCld} reduces examples like \Fig{nonconv}
to six ``signed'' intervals: four ``positive'': $[AD]$, 
$[AB]$, $[CD]$, $[BC]$ and two ``negative'': $[AC]$, $[BD]$. 

The convex case is revisited in \Sec{Conv} and \App{CalcDist}.
The construction of signed chord length distribution for
nonconvex body is described in \Sec{Noncon}. Some implications
to description of arbitrary bodies with nonuniform density
are briefly mentioned for completeness in \Sec{Nonunif} and \App{nonun}.
Other extensions, like polygonal trajectories are affected very
shortly in \Sec{path} and \App{pathlen}.

\section{Convex body}
\label{Sec:Conv}

\subsection{Basic geometrical models}
\label{Sec:GeoMod}

\begin{figure}[hbt]
\begin{center}
\includegraphics[scale=0.5]{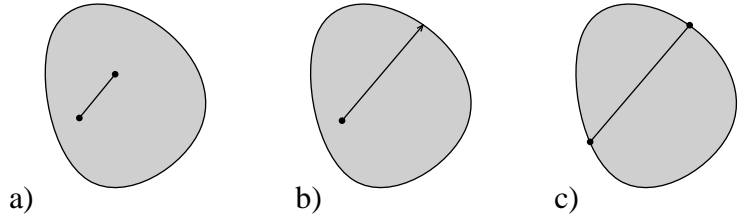}
\end{center}
\caption{Distributions: a) Distances between points.
b) Radii.
c) Chords.}\label{Fig:conv123}
\end{figure}

There are many different functions and relations between them used for
description properties of convex bodies %
\cite{Gil00,MRG03,MRD03,BR01,Str01,Han03,GMR05,Ken,San,Mat,Dir,Kel71,Maz03}.
In present paper are considered three different kinds of distributions: 
distances between points \Fig{conv123}a, lengths of
radii (segments) \Fig{conv123}b, and lengths of chords \Fig{conv123}c.
It may be useful to describe precisely models of generation of each distribution,
to avoid some problems with ambiguity, similar with widely known Bertrand 
paradox \cite{Ken,Prdx}. 
\begin{Defn}\label{defdst}
Distances distribution function is $\Dst(l) = \int_0^l\dst(x)dx$ with density $\dst(l)$.
The distances are defined by pairs of points inside of the body $\CB$.
Both points are from independent uniform distributions,
\Fig{conv123}a.
\end{Defn}
\begin{Defn}\label{defseg}
Radii distribution function is $\Seg(l) = \int_0^l\seg(x)dx$ with density $\seg(l)$.
The radii are defined as segments of rays from a point inside of the body $\CB$ to the 
surface. The points are from uniform distribution and directions of the rays are isotropic,
\Fig{conv123}b.
\end{Defn}
\begin{Defn}\label{defcld} 
Chord lengths distribution function is $\Cld(l) = \int_0^l\cld(x)dx$ with density $\cld(l)$.
The chords are defined by intersection of the body $\CB$ with isotropic uniform
distribution of lines, \Fig{conv123}c.
\end{Defn}

It is also convenient to use {\em autocorrelation function $\cor(l)$} related
with distances distribution for three-dimenional body as 
\begin{equation}
\dst(l) = 4 \pi l^2 \cor(l)/V,
\tag{\ref{cor2dst} in \App{CalcDist}}
\end{equation}
where $V$ is volume of $\CB$. The expression for autocorrelation function 
for body with arbitrary density \Eq{corr} together with derivation of 
\Eq{cor2dst} for constant density is recollected for completeness below 
in \App{AutoCor}.

There is remarkable correspondence between these densities 
\cite{Gil00,MRG03,MRD03,BR01,Str01,Han03,Kel71,Maz03}:
\begin{equation}
 \frac{1}{\av{l}}\cld(l) = -\seg'(l) = \cor''(l), 
\label{der12}
\end{equation}
where average chord length $\av{l}=\int_0^\infty{l\cld(l)dl}$ may be expressed 
via volume $V$ and surface area $S$ using a relation for three-dimensional 
convex body derived in XIX century by Cauchy, Czuber and rediscovered later
by Dirac {\em et al} \cite{Ken,Mat,Dir,Kel71}
\begin{equation}
 \av{l} = 4\frac{V}{S}.
\label{Cau}
\end{equation}

Distribution of lengths of radii \cite{Maz03} is also known as
{\em interior source randomness} \cite{Kel71}.
Relation between $\cld(l)$ and $\seg(l)$ in \Eq{der12} is often
represented in integral form \cite{Kel71,Maz03}
\[\seg(l)= \av{l}^{-1}\int_l^\infty\cld(x)dx = \av{l}^{-1}\bigl(1-\Cld(l)\bigr).\]
Proportionality between $\cld(l)$ and second derivative of autocorrelation function
$\cor(l)$ in \Eq{der12} is also well known and widely used \cite{Gil00,BR01,Str01,Han03}.

Here is convenient for completeness of presentation and further explanation of
nonconvex case to derive all equalities in \Eq{der12}, considering $\cld(l)$, 
$\seg(l)$, $\cor(l)$ and $\dst(l)$ as generalized functions.

The usual definition of {\em generalized function} \cite{KolmFom} is a {\em continuous
linear functional} $T(\phi)$ on a space of a {\em test functions} $\phi$. For an 
integrable function $\psi$ the functional $T_\psi$ is defined for a test function 
$\phi(x)$ as
\begin{equation}
  T_\psi(\phi) = \int_{-\infty}^\infty \psi(x) \phi(x)  dx. 
\label{Tpsi}
\end{equation}

The {\em generalized derivative} \cite{KolmFom} is defined as functional
\begin{equation}
T'(\phi) \equiv -T(\phi').
\label{GenDer}
\end{equation}
Such a definition ensures derivative of any
order for $T_\psi$ with arbitrary integrable function $\psi$ and it justifies
use of generalized functions and derivatives in \Eq{der12}.

\subsection{Dirac's method of chords}
\label{Sec:DirCh}

The Dirac's method of chords \cite{Dir} uses transition from six-dimensional 
integral over pair of points in some convex body $\CB$ to expressions with 
chord length distribution, {\eg} 
\begin{equation}
\DI{\CB}(\varphi) \equiv
\frac{1}{V}\int_\CB \int_\CB\frac{\varphi(R)}{4 \pi R^2} d\ve{r}\, d\ve{r'} =
\frac{S}{4V}\int_0^\infty\!\!\cld(x)%
\left(\int_0^x\!\!\!\int_0^p\!\!\varphi(r)dr dp\right) dx,
\label{IntDir}
\end{equation}
where $R = |\ve{r'}-\ve{r}|$ is distance between points, $d\ve{r}$ and $d\ve{r'}$
are two three-dimensional volume elements, $S$ is surface area, and $V$ is
volume of $\CB$. Such an equation was derived in Ref.~\cite{Dir} for particular
function $\varphi(R) = \exp(-\alpha R)$.

It is shown in \App{CalcDist}, that derivation of \Eq{IntDir} has direct 
connection with equalities \Eq{der12}. Here is only outlined basic results
and few steps of derivation \Eq{IntDir}.

\medskip

First, the integral $\DI{\CB}$ may be expressed via function of distances $\dst(l)$.
From \Eq{Int2V2dst} with \Eq{DI2FI} follows quite understanding relation
\begin{equation}
\frac{1}{V}\DI{\CB}(\varphi) =
\frac{1}{V^2}\int_\CB \int_\CB\frac{\varphi(R)}{4 \pi R^2} d\ve{r}\, d\ve{r'} =
 \int_0^\infty\!\!\frac{\varphi(x)}{4 \pi x^2}\dst(x)dx. 
\label{IntDirDst}
\end{equation}

Due to \Eq{Int2Vcor} with \Eq{DI2FI} right-hand side of \Eq{IntDirDst} may be 
rewritten using autocorrelation  function $\cor(l)$ \Eq{corr}
\begin{equation}
\DI{\CB}(\varphi) = \int_0^\infty\!\!\cor(x)\varphi(x)dx 
\label{IntDirCor}
\end{equation}

The six-dimensional integral $\DI{\CB}$ may be reduced to a four-dimensional one
by two steps \cite{Dir} revisited in details in \App{CalcDist}.
In accordance with \lemm{LemV2dst} in \App{DistDist}, 
\Eq{IntVSR} and \Eq{DI2FI} it may be expressed
\begin{equation}
  \DI{\CB}(\varphi) = 
 \frac{1}{4 \pi V}\int_\CB\!\!d\ve{r}\int d\Omega\!%
\int_0^{\Rd_{\max}}\!\! \varphi(R) dR,  
\label{IntDirVSR}
\end{equation}
where $\Rd_{\max}$ is length of radius for given point and direction, 
$d \ve{r}\, d\Omega$ is five-dimensional integration on all points 
and directions. 
Next, \lemm{LemVSseg} in \App{DistRad} let us rewrite \Eq{IntDirVSR}
using radii density function $\seg(l)$
\begin{equation}
 \DI{\CB}(\varphi) = 
 \int_0^\infty\!\!\seg(x)%
 \left(\int_0^x\!\!\varphi(r)dr \right) dx. 
\label{IntDirSeg}
\end{equation}

Finally, due to rather standard arguments \cite{Dir}, revisited in \App{DistChord},
it is possible to rewrite \Eq{IntDirVSR} using four-dimensional integration on space 
of lines $\T$  
\begin{equation}
 \DI{\CB}(\varphi) =
 \frac{1}{4 \pi V}\int\!d\T%
 \!\int_0^{\Lh_\mathrm{ch}}\!\!dp\int_0^p\!\!\varphi(r)dr, 
\label{IntDirT}
\end{equation}
where $d\T$ is canonical invariant measure on the space of lines and
$\Lh_\mathrm{ch}$ is length of chord for given line intersecting
body $\CB$. 

Now it is possible to rewrite \Eq{IntDirT} using
\lemm{Lem2V2T} from \App{DistChord} and \Eq{Int2V2T} to produce 
initial \Eq{IntDir}.

\medskip

These integrals also justify use generalized functions, because
may be associated with linear functionals on some test function $\varphi$.
The relations between integrals may be considered as transformations of
these functionals without necessity to indicate any particular $\varphi$
and it is quite reasonable, because main purpose of this paper is rather
discussion about geometrical distributions than about calculation of
some integrals.

\begin{subequations}\label{GenDist}
The \Eq{IntDirCor} may be rewritten using \Eq{Tpsi} for space of test functions
defined on some interval $0 \le x \le l_{\max}$  
\begin{equation}
 \DI{\CB} = T_\cor \equiv \cor.
\label{GenCor}
\end{equation}
The \Eq{IntDirSeg} may be rewritten due to definition of generalized derivative
\Eq{GenDer} 
\begin{equation}
 \DI{\CB}(\phi') = 
 \int_0^\infty\!\!\seg(x) \phi(x) dx \quad \Longrightarrow \quad
 \DI{\CB}' = - T_\seg \equiv - \seg.
\label{GenSeg}
\end{equation}
Finally, from \Eq{IntDir} follows
\begin{equation}
 \DI{\CB}(\phi'') = 
 \frac{S}{4V}\int_0^\infty\!\!\cld(x) \phi(x) dx \quad \Longrightarrow \quad
 \DI{\CB}'' = \frac{S}{4V} T_\cld \equiv \frac{S}{4V} \cld = \frac{1}{\av{l}} \cld.
\label{GenCld}
\end{equation}
\end{subequations}
The \Eq{GenDist} correspond to \Eq{der12} for generalized functions
and derivatives. 

\medskip

The derivation of integral \Eq{IntDirCor} with autocorrelation function $\cor(l)$
is quite straightforward and revisited in \App{AutoCor} \Eq{Int2Vcor}.
The Dirac expression \Eq{IntDir} is directly derived from \Eq{IntDirCor} via
two integration by parts if \Eq{der12} is true.
So if we could consider \Eq{der12} as a ``definition'' of a function $\cld(l)$
via second derivatives of $\cor(l)$ it ensures \Eq{IntDir}. 

Such a property was used in some works for definition of formal (generalized) chord length
distribution via \Eq{der12} for body with arbitrary shape and density \cite{Str01,Han03}.
 
\section{Nonconvex body}
\label{Sec:Noncon}

\subsection{Formal integration by parts}

Let us consider application of \Eq{IntDir} to some nonconvex body $\NB$.
The distances distribution and autocorrelation function is defined for
nonconvex body and so it is possible to write
\begin{subequations}\label{IntDirNB4}
\begin{eqnarray}
\DI{\NB}(\varphi) \equiv
\int_\NB \int_\NB\frac{\varphi(R)}{4 \pi V R^2} d\ve{r}\, d\ve{r'} 
 &=& V\int_0^\infty\!\!\frac{\varphi(x)}{4 \pi x^2}\dst(x)dx \label{IntDirDstNB} \\
 &=& \int_0^\infty\!\!\cor(x)\varphi(x)dx \label{IntDirCorNB} \\
 &=& -\int_0^\infty\!\!\cor'(x)%
 \left(\int_0^x\!\!\varphi(r)dr \right) dx \label{IntDirCorNB'} \\
 &=& \int_0^\infty\!\!\cor''(x)%
 \left(\int_0^x\!\!\!\int_0^p\!\!\varphi(r)dr\,dp\right) dx. \label{IntDirCorNB''}
\end{eqnarray}
\end{subequations}
Here \Eq{IntDirDstNB} and \Eq{IntDirCorNB} coincide with 
\Eq{IntDirDst} and \Eq{IntDirCor} in convex case respectively,
but \Eq{IntDirCorNB'} and \Eq{IntDirCorNB''} are produced by formal
integrations by parts and need for geometrical interpretation. 
Let us denote for nonconvex body the signed radii and chord densities represented via 
(normalized) first and second derivative as $\sigseg(l)$ and $\sigcld(l)$ respectively.

\subsection{Radii (signed) density function}
\label{Sec:SigSeg}

\begin{figure}[hbt]
\begin{center}
\includegraphics[scale=0.24]{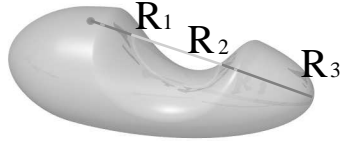}
\end{center}
\caption{Scheme of intervals for radii in nonconvex body}\label{Fig:NBrad}
\end{figure}

An analogue of \Eq{IntDirVSR} used for introduction of radii density
function has form
\begin{equation}
  \DI{\NB}(\varphi) = 
 \frac{1}{4 \pi V}\int_\NB\!\!d\ve{r}\int d\Omega\!%
\int_{\Rd \cap \NB}\!\! \varphi(R) dR,  
\label{IntDirVSRNB}
\end{equation}
where ${\Rd \cap \NB}$ is intersection of a body $\NB$ with a ray 
from a point inside the body \Fig{NBrad}.

It is possible to write
\begin{equation}
\int_{\Rd \cap \NB}\!\! \varphi(R) dR \equiv
 \sum_{j=0}^{\nI} \int_{R_{2j}}^{R_{2j+1}}\!\! \varphi(R) dR, \quad R_0 \equiv 0, 
\label{SumSegNB}
\end{equation}
where $\nI$ is amount of intervals $[R_{2j},R_{2j+1}]$ of ray
inside body $\NB$ with $R_k$ for $k > 0$ corresponding to $2 n_{\mathrm I}-1$ 
points of intersection of ray with surface of body \Fig{NBrad}.
The \Eq{SumSegNB} may be formally rewritten
\begin{equation}
\sum_{j=0}^{\nI} \int_{R_{2j}}^{R_{2j+1}}\!\! \varphi(R) dR = 
 \sum_{k=1}^{2 \nI-1}(-1)^{k-1}\int_0^{R_k}\!\! \varphi(R) dR, 
\label{pmSumSegNB}
\end{equation}

Let us denote $n_{\max}$ maximal possible amount of intervals for given
body and define $\seg_k(l)$, $k = 1,\ldots,2 n_{\max}-1$  as density
function for $k$-th interval length $R_k$. Then it is possible to
derive from \Eq{IntDirVSRNB} for 
\begin{equation}
\sigseg(l) = \sum_{k=1}^{2 n_{\max}-1}(-1)^{k-1}\seg_k(l)
\label{pmSeg}
\end{equation}
analogue of \Eq{IntDirSeg}
\begin{equation}
 \DI{\NB}(\varphi) = 
 \!\!\!\sum_{k=1}^{2 n_{\max}\!-1}\!\!\!(-1)^{k-1}\int_0^\infty\!\!\seg_k(x)%
 \left(\int_0^x\!\!\varphi(r)dr \right) dx =
 \int_0^\infty\!\!\sigseg(x)%
 \left(\int_0^x\!\!\varphi(r)dr \right) dx  
\label{IntDirSegNB}
\end{equation}

It is convenient to consider three distributions $\seg_1(l)$, $\seg_+(l)$, and 
$\seg_-(l)$, where
\begin{equation}
\seg_-(l) = \sum_{k=1}^{n_{\max}-1}\seg_{2k}(l), \quad
\seg_+(l) = \sum_{k=1}^{n_{\max}-1}\seg_{2k+1}(l), \quad
\sigseg(l) = \seg_1(l)+\seg_+(l)-\seg_-(l).
\label{seg123}
\end{equation}

Here ``positive'' and ``negative'' distributions, {\ie}
$\seg_1(l)+\seg_+(l)$ and $\seg_-(l)$ respectively may be nonzero
for the same $l$, {\eg} some radii $R_2$ may be equal to $R_3$
in other points. So \Eq{seg123} could not be considered as true 
Jordan decomposition defined as difference of two nonnegative 
functions with {\em nonoverlapping} support \cite{KolmFom}
\begin{equation}
\sigseg(l) = \seg^+(l)-\seg^-(l), 
\quad \forall l : \seg^+(l) \ge 0,~\seg^-(l) \ge 0,~\seg^+(l)\,\seg^-(l) = 0.
\label{JordSeg}
\end{equation}
In fact, for some nonconvex bodies $\sigseg(l) \ge 0$ 
and so $\seg^-(l)\equiv 0$ despite of nonzero $\seg_-(l)$ because 
of $\seg_1(l)+\seg_+(l) \ge \seg_-(l)$.

\medskip

Anyway, the scheme discussed above and expressed by \Fig{NBrad}, \Eq{pmSeg}, 
\Eq{IntDirSegNB}, and \Eq{seg123} provides a stochastic interpretation of 
$\sigseg(l)$. Similar with \defn{defseg} there is an uniform distribution of points 
inside a body and isotropic rays, considered as some kind of ``primary events'' 
for radius $R_1$ and $\seg_1(l)$. Each ray intersecting body $\NB$ more than
one time $\nI>1$ also produces two kinds of ``secondary'' events: $\nI-1$ 
radii $R_{2k+1}$ from a ``positive'' distribution $\seg_+(l)$ and $\nI-1$ radii 
$R_{2k}$ from a ``negative'' one $\seg_-(l)$.

Such a stochastic model also produces understanding description of some integrals
via averages, mathematical expectations {\em etc}., {\eg} 
\begin{equation}
 \int_0^\infty\seg_1(l)dl = \int_0^\infty\sigseg(l)dl = 1,
 \quad \int_0^\infty\seg_+(l)dl = \int_0^\infty\seg_-(l)dl,
\label{NormSeg}
\end{equation}
because each radius corresponds to $\nI$ ``positive'' events
and $\nI-1$ ``negative'' events, {\ie} contribution to ``total charge''
or ``balance'' of events $N \equiv N_+ - N_-$ is always one.

It is also useful to introduce distribution of total length of 
all segments for given radii $\seg_{\rm O}(l)$ (OSD, one-segment 
distribution), then
\begin{equation}
 \av{l}_\seg \equiv \int_0^\infty \sigseg(l)l\,dl = 
 \int_0^\infty \seg_{\rm O}(l)l\,dl,
\label{avOSD}
\end{equation}
because contribution for any ray  is
$\sum_{k=1}^{2 \nI-1}(-1)^{k-1}R_k = 
R_1 + \sum_{k=1}^{\nI-1}(R_{2k+1}-R_{2k})$, {\ie} total
length of all segments.

\subsection{Chord length (signed) density function}
\label{Sec:SigCld}

An analogue of integral \Eq{IntDirT} for nonconvex body $\NB$ may not
use continuous area of integration on $dp\,dr$, if chord intersects
$\NB$ along few intervals. It is similar with integration along set
of interval for radii \Eq{IntDirVSRNB} and \Eq{SumSegNB} in \Sec{SigSeg},
but decomposition is more complex because of two integrals. 

\begin{figure}[hbt]
\begin{center}
\includegraphics[scale=0.24]{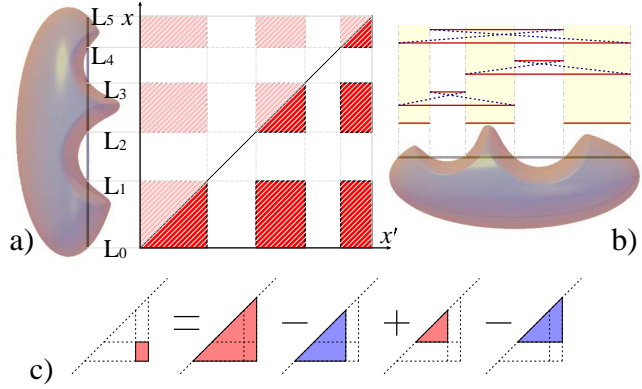}
\end{center}
\caption{a) Scheme of integration. b) Scheme of intervals
c) Decomposition of $\SQU_{j,k}$}\label{Fig:NBchord}
\end{figure}

For a chord with $\nI$ intervals there are $2 \nI$ points of
intersection ($L_0 \equiv 0, L_1$, $\ldots$, $L_{2k}, L_{2k+1}$, $\ldots$)
for $k=0, \ldots, \nI-1$. In \Eq{IntDirT} $p$ is coordinate along the chord 
and $r$ is distance between points. If to introduce new variables
$x=p$, $x'=p+r$, then for whole chord inside body
\begin{equation}
 \TRI^L(\varphi) \equiv
 \int_0^L\!\!\int_0^p\!\! \varphi(r)dr\,dp 
 = -\int_0^L\!\!\int_{0}^{x'}\!\!\varphi(x'-x) dx\,dx'
 = -\int_0^L\!\!\int_x^L\!\!\varphi(x'-x) dx' dx, 
\label{pr2xx'}
\end{equation}
but for few segments the integration of $\varphi(r) = \varphi(x'-x)$ on $dx\,dx'$ 
should include all points $x,x' \in \Lh \cap \NB$,  $x<x'$, where
$\Lh \cap \NB = \bigcup_{k=0}^{\nI-1} [L_{2k},L_{2k+1}]$ is union of
all intervals of given chord inside the body, \Fig{NBchord}a.
Here the minus signs in \Eq{pr2xx'} are due to term $-x$ in $\varphi(x'-x)$.

Let us denote $\LLint{\NB}$ area of integration described above and depicted on 
\Fig{NBchord}a as set of triangles and rectangles below of the diagonal $x=x'$
\begin{subequations}\label{AChr}
\begin{eqnarray}
\LLint{\NB} &=& \bigcup_{k=0}^{\nI-1}\!\!\TRI_k~{\textstyle\bigcup}
  \bigcup_{k=1,j=0}^{k<\nI,j < k}\!\!\!\SQU_{j,k} \,, 
\label{AChrL}\\
\TRI_k &=& \{(x,x') : L_{2k} \le x \le x' \le L_{2k+1} \}, 
\label{AChrT}\\
\SQU_{j,k} &=& \{(x,x') : L_{2j} \le x \le L_{2j+1},\,L_{2k} \le x' \le L_{2k+1}\}.
\label{AChrR}
\end{eqnarray}
\end{subequations}

Now it is possible to write analogue of \Eq{IntDirT} for nonconvex case
\begin{equation}
 \DI{\NB}(\varphi) = \frac{1}{4 \pi V}\int\!%
 \Bigl(-\!\iint_{\LLint{\NB}}\!\!\varphi(x'-x)dx\,dx'\Bigr) d\T.
\label{IntDirTxxNB}
\end{equation}

Due to \Eq{AChr} it may be rewritten
\begin{subequations}\label{SumCldNB}
\begin{multline}
-\!\iint_{\LLint{\NB}}\!\!\varphi(x'-x)dx\,dx' = \\
  = -\Bigl(\sum_{k=0}^{\nI-1}\!\iint_{\TRI_k}\!\!\varphi(x'-x)dx\,dx' +
\sum_{k=1}^{\nI-1}\sum_{j=0}^{k-1}\!\iint_{\SQU_{k,j}}\!\!\varphi(x'-x)dx\,dx'\Bigr), 
\label{SumCldInt}
\end{multline}
where
\begin{equation}
-\iint_{\TRI_k}\!\!\varphi(x'-x)dx\,dx' 
= \int_0^{L_{2k+1}-L_{2k}}\!\!\int_0^p\!\! \varphi(r)dr\,dp =
\TRI^{L_{2k+1}-L_{2k}}(\varphi)
\label{IntTRI}
\end{equation}
and integral on $\SQU_{j,k}$ may be decomposed on four terms illustrated on 
\Fig{NBchord}c
\begin{equation}
\begin{split}
-\iint_{\SQU_{j,k}}\!\!&\varphi(x'-x)dx\,dx' 
 = -\int_{L_{2j}}^{L_{2j+1}}\!\!\int_{L_{2k}}^{L_{2k+1}}\!\! \varphi(x'-x)dx' dx \\
&= \int_{L_{2j}}^{L_{2j+1}}\!\!\Bigl(\int_x^{L_{2k}}\!\! \varphi(x'-x)dx'-
 \int_x^{L_{2k+1}}\!\! \varphi(x'-x)dx'\Bigr)dx \\
&= \int_0^{L_{2j+1}}\!\!\int_x^{L_{2k}}\!\! \varphi(x'-x)dx' dx
  -\int_0^{L_{2j}}\!\!\int_x^{L_{2k}}\!\! \varphi(x'-x)dx' dx \\
  &\qquad -\int_0^{L_{2j+1}}\!\!\int_x^{L_{2k+1}}\!\! \varphi(x'-x)dx' dx
  +\int_0^{L_{2j}}\!\!\int_x^{L_{2k+1}}\!\! \varphi(x'-x)dx' dx\\
&=\TRI^{L_{2k}-L_{2j+1}}(\varphi) - \TRI^{L_{2k}-L_{2j}}(\varphi) 
 -\TRI^{L_{2k+1}-L_{2j+1}}(\varphi) + \TRI^{L_{2k+1}-L_{2j}}(\varphi)
\end{split}
\label{IntSQU}
\end{equation}
\end{subequations}

If $\nI$ is number of segments, there are $\nI$ (``positive'') terms 
in \Eq{IntTRI} and $4 \frac{\nI(\nI-1)}{2}$ terms ($\nI^2-\nI$ 
``positive'' and $\nI^2-\nI$ ``negative'') in \Eq{IntSQU}.

It is possible to decompose \Eq{IntDirTxxNB} using \Eq{SumCldNB}
\begin{multline}
 \DI{\NB}(\varphi) = \frac{1}{4 \pi V}\int\!%
 \Bigl(\sum_{k=0}^{\nI-1}\!\TRI^{L_{2k+1}-L_{2k}}(\varphi)\\
 + \sum_{k=1}^{\nI-1}\sum_{j=0}^{k-1}
 \bigl[\TRI^{L_{2k}-L_{2j+1}}(\varphi) + \TRI^{L_{2k+1}-L_{2j}}(\varphi)\bigr]\\
 - \sum_{k=1}^{\nI-1}\sum_{j=0}^{k-1}
 \bigl[\TRI^{L_{2k}-L_{2j}}(\varphi) +\TRI^{L_{2k+1}-L_{2j+1}}(\varphi)\bigr]
\Bigr)d\T.
\label{SumIntDirT}
\end{multline}
Let us recall \Eq{IntDirT} for convex body expressed using \Eq{IntTRI}
\begin{equation}
\DI{\CB}(\varphi) = \frac{1}{4 \pi V}\int\!\TRI^{\Lh_\mathrm{ch}}(\varphi)d\T.
\label{IntDirTRI}
\end{equation}
The \Eq{SumIntDirT} with three terms corresponds to sum of $2\nI^2-\nI$ integrals 
\Eq{IntDirTRI} arranged in three groups.
The first one includes all $\nI$ segments $[L_{2k},L_{2k+1}]$ of given chord.
Second one takes into account $\nI(\nI-1)/2$ pairs of intervals 
$[L_{2j},L_{2k+1}]$ and $[L_{2j+1},L_{2k}]$. Third ``negative'' term is for 
$\nI(\nI-1)/2$ pairs of intervals $[L_{2j},L_{2k}]$ and $[L_{2j+1},L_{2k+1}]$. 

For a chord with $\nI$ intervals there are $\nI^2$ ``positive'' and 
$\nI^2-\nI$  ``negative'' terms. 
The scheme is depicted on \Fig{NBchord}b for $\nI=3$, 
there ``negaitive'' intervals are drawn by dashed lines.

It is possible to decompose $\sigcld(l)$ on three parts corresponding terms
in \Eq{IntDirTRI}: $\cld_1(l)$, $\cld_+(l)$, $\cld_-(l)$.
Here the $\cld_1(l)$ up to normalizing multiplier corresponds to density
$\cld_{\rm M}$ of multi-chord distribution (MCD) discussed in \Sec{Intro}.
\begin{equation}
 \cld_1(l) = c_{\rm M} \cld_{\rm M}(l), \quad
 c_{\rm M} = \int_0^\infty \cld_1(l) dl.
\label{cld1M}
\end{equation}
If normalization of $\sigcld(l)$ is required, it should be expressed
\begin{equation}
\sigcld(l) = c_{\rm M}^{-1}[\cld_1(l)+\cld_+(l)-\cld_-(l)]
= \cld_{\rm M}(l)+c_{\rm M}^{-1}[\cld_+(l)-\cld_-(l)].
\label{cld123}
\end{equation}

Here again ``positive'' $c_{\rm M}^{-1}[\cld_1(l)+\cld_+(l)]$ and 
``negative'' $c_{\rm M}^{-1}\cld_-(l)$ terms of \Eq{cld123} can be
overlapped and so formally could not be considered as Jordan 
decomposition \cite{KolmFom}
\begin{equation}
\sigcld(l) = \cld^+(l)-\cld^-(l), 
\quad \forall l : \cld^+(l) \ge 0,~\cld^-(l) \ge 0,~\cld^+(l)\,\cld^-(l) = 0.
\label{JordCld}
\end{equation}

\medskip

Let us write finally analogue of \Eq{IntDir} for nonconvex case
\begin{equation}
\DI{\NB}(\varphi) \equiv
\frac{1}{V}\int_\NB \int_\NB\frac{\varphi(R)}{4 \pi R^2} d\ve{r}\, d\ve{r'} =
\ell_{\NB}^{-1}\int_0^\infty\!\!\sigcld(x)%
\left(\int_0^x\!\!\!\int_0^p\!\!\varphi(r)dr dp\right) dx,
\label{IntDirNB}
\end{equation}
where $\ell_{\NB}$ is some constant. It was shown earlier for convex body 
$\ell_{\CB} = \av{l} = 4V/S$.

Definition of {\em signed chord distribution} as quantity proportional to 
second derivative of autocorrelation function suggested in \Sec{Intro}
together with comparison of \Eq{IntDirNB} and \Eq{IntDirCorNB''} 
produces rather formal equations like
\begin{subequations}\label{ellNB}
\begin{equation}
\ell_{\NB}^{-1}\sigcld(l) = \cor''(l) = -\sigseg'(l), \qquad 
\ell_{\NB}^{-1} = \int_0^\infty \cor''(l) dl = \cor'(0),
\label{ellNBcor'}
\end{equation}
but \Eq{cld123} also describes ``formal'' $\sigcld(l)$ via geometrical
distributions of segments of chord in nonconvex body. 

It is also possible to write analogue \Eq{avlcld}
\begin{equation}
\ell_{\NB} = \ell_{\NB}\int_0^\infty \sigseg(l) dl = 
-\ell_{\NB}\int_0^\infty \sigseg'(l) l\,dl
= \int_0^\infty \sigcld(l) l\,dl = \av{l}.
\label{ellNBavl}
\end{equation}
If to use \Eq{IntDirNB} with $\varphi(l)=4\pi l^2$ it is possible to derive from
\Eq{IntDirCorNB''} yet another one relation for an average
\begin{equation}
\int_\NB \int_\NB\frac{d\ve{r}\, d\ve{r'}}{V} =
\frac{V^2}{V}=\ell_\NB^{-1}\int_0^{\infty}\sigcld(l)\frac{4\pi l^4}{12}dl
\quad\Longrightarrow\quad
\ell_\NB = \frac{\pi \av{l^4}}{3 V} 
\label{ellNBavl4}
\end{equation}
\end{subequations}

\medskip

It is possible to use stochastic model to clarify some expressions
above and derive other useful results.
There is uniform isotropic distribution of lines described in \defn{defcld}
and corresponding to ``primary'' events. If such line intersects body
$n_I$ times, there are $2\nI^2-\nI$ ``secondary'' events. There are
$n_I$ segments of given chord together with $\nI(\nI-1)$ ``positive'' and 
``negative'' intervals described above.

\paragraph*{Note on event counting:}\label{totch}
An essential difference with radii distribution is contribution $\nI \ge 1$ for 
each chord to ``total charge'' $N = N_+ -N_-$.
So number of lines $N_l$ is not equal with $N$ and $N_l/N \to c_{\rm M}$ for
$N \to \infty$. The similar effect is true for work with usual MCD case and so 
should not be considered as specific difficulty of signed chord distribution.

It is also possible to express $c_{\rm M}$ using $\cld_{\rm O}(l)$ density
for one-chord case (OCD) also mentioned in \Sec{Intro}, because contribution 
to average length for $\cld_1$ (without normalization on $c_{\rm M}$) 
is sum of all segments and it is the same for OCD case
\begin{equation}
 \int_0^\infty\!\!\!\cld_{\rm O}(l) l\, dl = \int_0^\infty\!\!\!\cld_1(l) l\, dl =
 c_{\rm M}\!\int_0^\infty\!\!\!\cld_{\rm M}(l) l\, dl  =
 c_{\rm M}\!\int_0^\infty\!\!\!\sigcld(l) l\, dl,
 \quad c_{\rm M} 
 =\frac{\av{l}_{\rm O}}{\av{l}}.
\label{cmlol}
\end{equation}

There is also quite similar expression for $\av{l^2}$. The contribution to
$l^2$ for one chord due to \Eq{SumIntDirT}
\begin{multline}
 \sum_{k=0}^{\nI-1}(L_{2k+1}-L_{2k})^2\\
 + \sum_{k=1}^{\nI-1}\sum_{j=0}^{k-1}\bigl(
 (L_{2k}-L_{2j+1})^2 + (L_{2k+1}-L_{2j})^2
 - (L_{2k}-L_{2j})^2 - (L_{2k+1}-L_{2j+1})^2 \bigr) \\
 = \sum_{k=0}^{\nI-1}(L_{2k+1}-L_{2k})^2
 + 2 \sum_{k=1}^{\nI-1}\sum_{j=0}^{k-1} (L_{2k+1}-L_{2k})(L_{2j+1}-L_{2j}) \\
 =\Bigl(\sum_{k=0}^{\nI-1}(L_{2k+1}-L_{2k})\Bigr)^2
\label{pmSumSq}
\end{multline}
Due to \Eq{pmSumSq} this contribution (to $\cld_1+\cld_+-\cld_-$) is always 
equal to square of sum of all segments, {\ie} coincides with OCD case.
So, due to \Eq{pmSumSq} and \Eq{cld123}
\begin{equation}
 \int_0^\infty\!\!\!\cld_{\rm O}(l) l^2 dl = 
 \int_0^\infty\!\![\cld_1(l)+\cld_+(l)-\cld_-(l)]\,l^2 dl = 
 c_{\rm M}\!\int_0^\infty\!\!\!\sigcld(l) l^2 dl,
 \quad 
 c_{\rm M}  = \frac{\av{l^2}_{\rm O}}{\av{l^2}}.
\label{cmlol2}
\end{equation}

Finally, let us recall few methods for calculation of multiplier $\ell_\NB$.  
It is discussed in Ref.~\cite{GMR05} that for convex and nonconvex
case $\cor'(0) = S/4V$ and so due to \Eq{ellNBcor'}, it is possible
to use $\ell_{\NB} =\cor'(0)^{-1} = 4V/S$ similar with nonconvex case. 
There are also interesting results \cite{Gil00,MRG03} concerning $\av{l}_{\rm O}$ 
and $\av{l}_{\rm M}$ for nonconvex body. 
It was proven for wide class of nonconvex bodies \cite{MRG03} $\av{l}_{\rm M} = 4V/S$,
and so $\av{l} = \av{l}_{\rm M} = 4V/S$ and \Eq{ellNBavl} again ensures 
$\ell_{\NB} = \av{l} = 4V/S$.

It was also mentioned \cite{MRG03}
yet another useful result $\av{l}_{\rm O} = 4V/S^*$, where $S^*$ is surface area
of convex hull and so due to \Eq{cmlol} $c_{\rm M} = S/S^*$. 
On the other hand, it is not quite clear, if equations with $S^*$ 
are true for nonconvex body that may not be represented
as convex body with few convex holes. 

\smallskip

The stochastic model considered here also clarify use Monte-Carlo methods for
calculation of signed chord length distribution used in \Eq{IntDirNB}. 
It is necessary to generate uniform isotropic random lines using methods 
discussed elsewhere \cite{MCbund}. For each chord, intersecting a body
$\nI$ times, lengths of all $2\nI^2-\nI$ segments represented in 
\Eq{SumIntDirT} should be taken into account with proper signs. 

It is possible to generate either one signed distribution or
two distributions for ``positive'' and ``negative'' segments to 
represent $\sigcld(l)$ via difference like \Eq{cld123}. 
It is also necessary to use proper normalization, {\ie} use the ``total
charge'' $N = N_+ - N_-$, rather than number of lines $N_l$ 
(see {\em Note on event counting} above on page \pageref{totch}).

\section{Nonuniform case}
\label{Sec:Nonunif}

The nonuniform case has more ambiguity in definitions and difficulty with geometrical 
interpretation. It is only briefly mentioned here. Two analogues of \Eq{IntDir} are 
considered in \App{nonun} and have form
\begin{equation}
 \iint\rho(\ve{r}) \rho(\ve{r'})%
 \frac{\varphi(\Delta_{\ve{r},\ve{r}'})}{4 \pi |\ve{r}-\ve{r}'|^2}
 d\ve{r}\, d\ve{r'} =
 C_\vartriangle\int_0^\infty\!\!\cld_\vartriangle(x)%
 \left(\int_0^x\!\!\!\int_0^p\!\!\varphi(r)dr\,dp\right) dx,
\label{IntDirNonUn}
\end{equation}
where $\Delta_{\ve{r},\ve{r}'}$ is some function, $C_\vartriangle$ is constant
and $\cld_\vartriangle(x)$ is a density for chosen function $\Delta_{\ve{r},\ve{r}'}$. 

A simple choice $\Delta_{\ve{r},\ve{r}'}=|\ve{r}-\ve{r}'|$ is discussed
in \App{nonun_dist} \Eq{Int2Mcor} and it is proven the correspondence 
with definition of CLD $\acute{\cld}(l) \hookrightarrow \cld_\vartriangle(l)$ 
via second derivative \cite{Str01,Han03}.

In some applications it is more appropriate to consider ``optical'' length
$\Delta_{\ve{r},\ve{r}'}=\opt{\ve{r}}{\ve{r}'}$ defined by \Eq{OptLen} as an 
amount of substance along straight line between $\ve{r}$ and $\ve{r}'$.
For uniform case and convex body it is proportional to distance between points, 
but for nonuniform case it is another integral \Eq{FJ} reproduced in \App{nonun_opt}.

For such an integral there is also analogue of \Eq{IntDir}. It is expression 
\Eq{Int2Mocld} with distribution of ``optical'' chord lengths 
$\tilde\cld(l) \hookrightarrow \cld_\vartriangle(l)$ 
used in \lemm{Lem2V2ocld} in \App{nonun_opt}.
The complete proof of \Eq{Int2Mocld} with formal ``cancellation'' of two terms
with $\rho$ may be found in \App{nonun_opt}. This case may be considered as even 
more direct analogue of convex body with uniform density \Eq{IntDir}, because 
$\tilde\cld(l)$ is certainly nonnegative due to quite clear geometrical definition.

The $\tilde\cld(l)$ is also may be used for nonconvex body $\NB$ with uniform 
density $\rho=1$ by formal consideration of union of given body and complement
to convex hull with $\rho=0$. In such a case ``optical'' length of chord
corresponds to sum of all intervals inside $\NB$, {\ie} definition of 
OCD case \cite{Gil00,MRG03,MRD03} mentioned in \Sec{Intro}. 

The application of OCD case for nonconvex body with uniform density was 
suggested already in initial paper about method of chords \cite{Dir}.

\section{Applications to arbitrary paths}
\label{Sec:path}

A natural extension of problems discussed in present paper --- is the
consideration of polygonal paths. There are interesting results, concerning
trajectories of random walk, {\eg} validity of Cauchy formula \Eq{Cau} 
for average path length inside the body \cite{BF03,Maz4J,Maz4E,Maz05,BCM05}.
\App{pathlen} contains a rather illustrative geometrical derivation of this 
result for uniform isotropic case.

Possibility of application to some tasks concerning arbitrary trajectories
is important property of method of chords both for convex \Eq{IntDir} and 
nonconvex case \Eq{IntDirNB}. 
Sometimes consideration of such integrals uses understanding, but simplified 
models with propagation of particles along straight lines.
Such a suggestion is not always justified, because in most cases considered
above the segments, radii and chords --- are formal subspaces of integration.
The only necessary condition is isotropy and uniformity of model, justifying
use of function $\varphi(|\ve{r}'-\ve{r}|)$, instead of some general function 
$\varphi(\ve{r}',\ve{r})$.

Formally, the only example of more general function mentioned here is 
$\varphi(\opt{\ve{r}'}{\ve{r}})$ used in \Eq{Int2Mocld} and discussed
in \App{nonun_opt}. For such a model consideration with propagation
along line is justified due to definition of $\opt{\ve{r}'}{\ve{r}}$
in \Sec{Nonunif} and \App{nonun_opt} \Eq{OptLen}.

\begin{figure}[hbt]
\begin{center}
\includegraphics[scale=0.24]{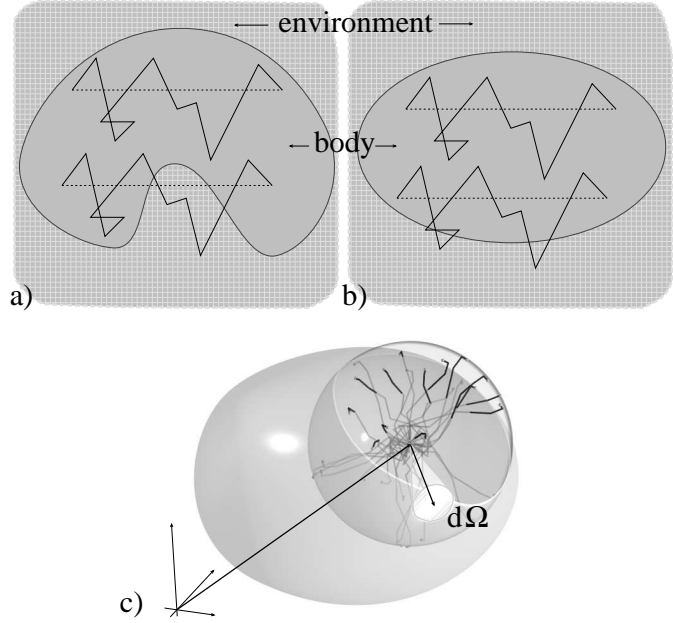}
\end{center}
\caption{Trajectories and lines: a) Nonconvex body b) Convex body
c) Spherical symmetry in distributions of trajectories}\label{Fig:ncbpaths}
\end{figure}

Here should be compared two cases. One model with propagation along
lines makes possible to use method of chord for convex, nonconvex
and even nonuniform case \Sec{Nonunif} and \App{nonun_opt}. Other model
may be used only for uniform isotropic case, but it is possible to consider
also propagation with scattering, that may be described
by some function $\varphi(|\ve{r}'-\ve{r}|)$ due to spherical and 
translational symmetry \Fig{ncbpaths}. 

The basic theme of presented paper is the second model described
by signed density function $\sigcld(l)$. The first model was discussed
for completeness in \Sec{Nonunif} and \App{nonun_opt} and it is shown,
that it always described by some nonnegative density function $\tilde\cld(l)$
without essential difference in construction for nonconvex and even
nonuniform case.
The construction of signed chord density function is much complicated
and was described in \Sec{SigCld}.

Let us consider examples of application of both cases. On \Fig{ncbpaths}
is represented propagation along some trajectories. For nonconvex case
\Fig{ncbpaths}a any trajectory may intersect body more than once. 
For straight line it is always possible to use first model with $\tilde\cld(l)$.
For nonconvex case and environment with neglectful density $\opt{\ve{r}'}{\ve{r}}$ 
is simply sum of all segments inside body along given line.

A polygonal trajectory may intersect body more than once for both
nonconvex and convex cases \Fig{ncbpaths}. If body and environment 
is the same uniform isotropic media, it is possible to use integrals
with $\varphi(|\ve{r}'-\ve{r}|)$, otherwise there are ``boundary effects,''
because different trajectories intersect environment by different ways
depending on concrete positions $\ve{r}'$ and $\ve{r}$ even for convex
case \Fig{ncbpaths}b. 

\smallskip

An illustrative example of an application is a set of random trajectories 
from uniformly distributed points $\ve{r}$ inside the body with isotropic 
initial directions (see \Fig{ncbpaths}c) and function 
$\varphi(R) = \varphi(|\ve{r}'-\ve{r}|)$, treated 
as probability density of cancellation of trajectory (absorption) on distance $R$
from an origin $\ve{r}$. Then \Eq{IntDir} or \Eq{IntDirNB} describe probability for 
a trajectory of such random walk to be finished inside the body. 

\newpage

\appendix

\section{Calculation of distributions for convex body}
\label{App:CalcDist}

\subsection{Distribution of distances}
\label{App:DistDist}

Let us consider some convex body $\CB \subset \R^3$, function
$\Phi(x)$ and integral 
\begin{equation}
 \FI{\CB}(\Phi) = 
 \frac{1}{V^2}\int_\CB \int_\CB \Phi\bigl(|\ve{r}'-\ve{r}|\bigr) d\ve{r}\, d\ve{r}',
\quad \ve{r} = (x,y,z),\quad d\ve{r} = dx\,dy\,dz,
\label{Int2V}
\end{equation}
where $V$ is volume of $\CB$.
It is six-dimensional integral with the function $\Phi$ of a distance $|\ve{r}'-\ve{r}|$ 
for pair of points $\ve{r},\ve{r}' \in \CB$. 
The multiplier $1/V^2$ is used for normalization $\FI{\CB}(1) = 1$ and
\Eq{Int2V} may be also treated as an average of the function $\Phi$
of one variable and expressed via single integration due to lemma below.

\begin{Lemm}\label{LemV2dst}
The linear functional $\FI{\CB}(\Phi)$ 
defined by \Eq{Int2V} may be rewritten
\begin{equation}
 \FI{\CB}(\Phi) = \int_0^\infty \Phi(x) \dst(x) dx,
\label{Int2V2dst}
\end{equation}
where $\dst(x)$ is density of distances distribution in body $\CB$ introduced
in \defn{defdst}.
\end{Lemm}

\begin{proof}
The linear functional $\FI{\CB}(\Phi)$ also should be considered 
as {\em a generalized function} and for regular case it may be
represented via integral like \Eq{Tpsi}, {\ie} \Eq{Int2V2dst} with some 
$\dst(x)$. Let us show, that $\dst(x)$ is density for distribution of distances.

It is enough to consider $\FI{\CB}$ with functions $\Phi(x)= \Theta(l-x) \equiv 
\Theta_l^\neg(x)$, there $\Theta$ is Heavyside step function, {\ie} $\Theta_l^\neg(x)$
is equal to zero for $x > l$ and unit otherwise. For such a function $\FI{\CB}$
integrates over all pairs of points in $\CB \times \CB$ with distances less than 
$l$ and due to \Eq{Int2V2dst} it is possible to write equation for probability 
$\FI{\CB}(\Theta_l^\neg) = 
\Prob(|\ve{r}'-\ve{r}| < l) = \int_0^l \dst(x) dx \equiv \Dst(l)$ 
and it coincides with definition of density $\dst(x)$ for distances distribution 
function $\Dst(x)$.

For extension of the proof for non-regular case, it is possible to
use directly the distances distribution function $\Dst(x)$ and to 
write instead of \Eq{Int2V2dst} Lebesgue-Stieltjes integral \cite{KolmFom,Korn}
\begin{equation}
 \FI{\CB}(\Phi) = \int_0^\infty \Phi(x) d F_\dst(x).
\label{Int2V2dstLS}
\end{equation}
\end{proof}

For consideration of both regular and non-regular cases it is
also possible to treat $\dst$ as a generalized function.
In such a case representation of $\FI{\CB}(\Phi)$ as an integral \Eq{Int2V2dst}
may look as a not very rigor representation of a tautology like 
$\FI{\CB} = \dst \equiv T_\dst$, because a generalized function by definition is a 
linear functional \cite{KolmFom}. 

\smallskip

The functional $\FI{\CB}$ is connected with $\DI{\CB}$ defined in
\Sec{DirCh}, \Eq{IntDir} by straightforward relations
\begin{equation}
 \DI{\CB}(\varphi) = V\FI{\CB}\bigl(\frac{\varphi(x)}{4 \pi x^2}\bigr), 
 \quad \FI{\CB}(\Phi) = \frac{1}{V}\DI{\CB}\bigl(4 \pi x^2 \Phi(x) \bigr),
\label{DI2FI}
\end{equation}
produced by the choise 
$\displaystyle\Phi(x) = V\frac{\varphi(x)}{4 \pi x^2}$.

\smallskip

The meaningfulness \lemm{LemV2dst} is not only integral 
representation \Eq{Int2V2dst}, but association of $\FI{\CB}$ with 
distribution of distances. It is simple demonstration of some methods 
to avoid Bertrand-like paradoxes, because $\dst(x)$ is not only possible 
density for distribution of distances between points in a body.

For calculation of $\FI{\CB}$ 
\Eq{Int2V} is used integration on six-dimensional space with natural
Euclidean measure on $\R^6 = \R^3 \times \R^3$ and it is in agreement
with {\em independent uniform distributions of both points} used
in \defn{defdst}.  

An example of alternative measure was used in Ref.~\cite{Maz03}: it 
was considered distribution of chords like in \defn{defcld}
above and segment of line between pair of points on the chords. 
A distribution function for lengths of the segments, {\ie} distances between 
ending points may be also calculated using Heavyside step function \cite{Maz03}, 
but density is not equivalent with $\dst(x)$. 

Yet another example of similar measure may be constructed if we 
consider first point $\ve{r}$ from uniform distribution, but the 
second one $\ve{r}'$ is generated with isotropic distribution of relative
directions $\ve{R} \equiv \ve{r}'-\ve{r}$ and uniform distribution of distances 
$|\ve{R}| \equiv |\ve{r}'-\ve{r}|$ between points.
In such a case  distributions of points $\ve{r}$ and $\ve{r}'$ are correlated
and relatively to natural Euclidean measure on 
$(\ve{r},\ve{r}') \in \R^3 \times \R^3 = \R^6 \supset \CB \times \CB$ here is
necessary to introduce multiplier $|\ve{R}|^{-2}=|\ve{r}'-\ve{r}|^{-2}$.

Up to constant normalizing multiplier the alternative density of distances 
introduced above may be expressed as $\dst(x)/x^2$, {\ie} proportional to
autocorrelation function \cite{Gil00,BR01,Str01,Han03} discussed further.

\subsection{Autocorrelation function}
\label{App:AutoCor}

For body with density $\rho(\ve{r})$, $\ve{r} \in \R^3$ is defined the 
autocorrelation function $\cor(\ve{r})$, $\ve{r} \in \R^3$ or $\cor(l)$, $l \in \R$
\begin{equation}
 \cor(\ve{r}) = \int_{\R^3} \rho(\ve{r}') \rho(\ve{r}+\ve{r}') d \ve{r}', \quad
 \cor(l) = \frac{1}{4\pi l^2}\int_{\Sph_l} \cor(\ve{r}) d\Omega, \quad 
 d\Omega = \sin\theta\, d\theta\, d\phi,
\label{corr}
\end{equation}
{\ie} $\cor(l)$ is an average of $\cor(\ve{r})$ on sphere with radius $l$,
$\{\Sph_l : |\ve{r}|=l \}$. 

In simplest case of constant density 
$\rho(\ve{r}) = 1$ for $\ve{r} \in \CB$ and zero otherwise.
For such a case $\cor(0) = V$, and it is convenient to rescale 
$\varrho(\ve{r}) \equiv \rho(\ve{r})/\sqrt{V}$ for normalization $\cor(0) = 1$. 
It is possible to rewrite \Eq{Int2V}
\begin{eqnarray}
\FI{\CB}(\Phi) &=&  
 \frac{1}{V} \int_{\R^3} \int_{\R^3} \varrho(\ve{r}) \varrho(\ve{r}')%
 \Phi\bigl(|\ve{r}'-\ve{r}|\bigr) d\ve{r}\, d\ve{r}' \nonumber \\
 &=& \frac{1}{V} \int_{\R^3} \int_{\R^3} \varrho(\ve{r}) \varrho(\ve{r+R})%
 \Phi\bigl(|\ve{R}|\bigr) d\ve{r}\, d\ve{R} \qquad (\ve{R=r'-r}) \nonumber \\
 &=& \frac{1}{V} \int_{\R^3} \cor(\ve{R})\Phi\bigl(|\ve{R}|\bigr) d\ve{R} 
\label{Int2Vcorr} \\
 &=& \frac{4\pi}{V} \int_0^\infty l^2 \cor(l)\Phi(l) dl,
\label{Int2Vcor}
\end{eqnarray}
where \Eq{Int2Vcor} is produced from \Eq{Int2Vcorr} by integration 
over spheres $\Sph_l$.

Comparison of \Eq{Int2Vcor} and \Eq{Int2V2dst} with arbitrary function $\Phi(l)$
produces relation between $\cor(l)$ and $\dst(l)$
\begin{equation}
\dst(l) = \frac{4 \pi}{V} l^2 \cor(l)
\label{cor2dst}
\end{equation}
already mentioned earlier in \Sec{GeoMod}.

\subsection{Distribution of radii}
\label{App:DistRad}

\begin{figure}[hbt]
\begin{center}
\includegraphics[scale=0.24]{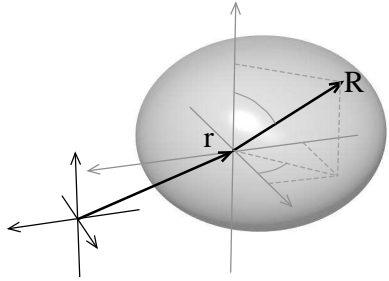}
\end{center}
\caption{Scheme of integration along radii with length
$\Rd(\ve{r},\theta,\phi)=|\ve{R}|$}\label{Fig:seg}
\end{figure}

Let us introduce new variable $\ve{R}=\ve{r}'-\ve{r}$ in \Eq{Int2V}
and rewrite integral on $d\ve{R}$ using spherical coordinates
\begin{eqnarray}
 \FI{\CB}(\Phi) &=& \frac{1}{V^2}\int_\CB d\ve{r} 
 \int_0^\pi \sin\theta d\theta\int_0^{2\pi}d\phi 
 \!\int_0^{\Rd(\ve{r},\theta,\phi)}\!\! R^2 \Phi(R) dR \nonumber \\
 &=& \frac{1}{V^2}\int_\CB d\ve{r} 
 \int_\Sph d\Omega 
 \!\int_0^{\Rd(\ve{r},\ve{\Omega})}\!\! R^2 \Phi(R) dR,  
\label{IntVSR}
\end{eqnarray}
where $R^2 \sin \theta$ is Jacobian in spherical coordinates
($R$, $\theta$, $\phi$) of vector $\ve{r}'-\ve{r}$, $\Sph$ is unit
sphere, and $\Rd(\ve{r},\theta,\phi)=\Rd(\ve{r},\ve{\Omega})$ is length of radius 
(segment) mentioned in \defn{defseg}, {\ie} distance from
point $\ve{r}$ to surface of body in direction represented by
spherical angles $\theta,\phi$ or unit vector $\ve{\Omega} \in \Sph$,
\Fig{seg}. 

Let us introduce linear functional
\begin{equation}
 \FII{\CB}(\Psi) =
 \frac{1}{4\pi V}\int_\CB\int_\Sph \Psi[\Rd(\ve{r},\ve{\Omega})] d\ve{r}  d\Omega, 
\label{IntVS}
\end{equation}
where $\ve{r} \in \CB \subset \R^3$, $\ve{\Omega} \in \Sph$, and
$\Rd(\ve{r},\ve{\Omega})$ is notation for length of radius introduced in \Eq{IntVSR}.
Here $\FII{\CB}(1)=1$ and $4\pi V$ is normalization, {\ie} 5D volume of set 
$\CB \times \Sph$ with respect to canonical measure on $\R^3 \times \Sph$.  
It is analogue of
normalization of \Eq{Int2V} with $V^2$, {\ie} 6D volume of set $\CB \times \CB$ 
with respect to canonical measure on $\R^3 \times \R^3$.

\begin{Lemm}\label{LemVSseg}
The linear functional $\FII{\CB}(\Phi)$ 
defined by \Eq{IntVS} may be rewritten
\begin{equation}
\FII{\CB}(\Psi) = \int_0^\infty \Psi(x) \seg(x) dx 
\label{IntVS2seg}
\end{equation}
where $\seg(x)$ is density of radii distribution in body $\CB$ introduced
in \defn{defseg}.
\end{Lemm}
\begin{proof}
The proof is similar with \lemm{LemV2dst}. The \Eq{IntVS2seg} is again almost 
tautology for generalized functions $\FII{\CB} = \seg \equiv T_\seg$. Let us anew
consider $\FII{\CB}$ with Heavyside step function $\Theta_l^\neg(x)$, to integrate 
over points in $\CB \times \Sph$ associated with radii less than given $l$.

The measures of integration in $\FII{\CB}$ correspond to uniform distribution of
points $\ve{r}$  and isotropic distribution of directions $\ve{\Omega}$ on unit 
spheres (due to transition to spherical coordinates in second integral).
It is in agreement with \defn{defseg} and so 
$\FII{\CB}(\Theta_l^\neg) = \Prob(\Rd(\ve{r},\ve{\Omega}) < l) 
= \int_0^l \seg(x) dx \equiv \Seg(l)$ and meets 
definition of density $\seg(x)$ for radii length distribution 
function $\Seg(x)$.
\end{proof}

The functional $\FII{\CB}$ \Eq{IntVS} let us simplify \Eq{IntVSR}
\begin{equation}
 \FI{\CB}(\Phi) = 
 \FII{\CB}\left(\frac{4 \pi}{V}\int_0^{x}\!\! R^2 \Phi(R) dR\right)
\label{IntVSR2V2}
\end{equation}
and substition of \Eq{IntVSR2V2} to \Eq{IntVS2seg} produces yet another
expression for $\FI{\CB}(\Phi)$ 
\begin{equation}
 \FI{\CB}(\Phi) = \int_0^\infty%
 \!\!\left(\frac{4 \pi}{V}\int_0^x R^2\Phi(R) dR\right) \seg(x) dx.
\label{Int2V2seg}
\end{equation}

\subsection{Autocorrelation function and distribution of radii}

Due to definition \Eq{cor2dst} it is possible to rewrite \Eq{Int2V2dst}
\begin{equation}
 \FI{\CB}(\Phi) = \int_0^\infty \Phi(x) \frac{4 \pi}{V} x^2 \cor(x) dx.
\label{Int2cor}
\end{equation}

Yet another expression
\begin{equation}
 \FI{\CB}(\Phi) = 
 -\int_0^\infty \!\!\left(\frac{4 \pi}{V}\int_0^x R^2\Phi(R) dR\right) \cor'(x) dx,
\label{Int2cor2}
\end{equation}
may be produced from \Eq{Int2cor} using integration by parts. 
The equality of \Eq{Int2cor2} and \Eq{Int2V2seg} corresponds
to second equation in \Eq{der12}
\begin{equation}
 -\seg(x) = \cor'(x).
\label{der1}
\end{equation}
The integral representations above let us also consider \Eq{der1} as 
derivative of generalized function.

\subsection{Chord length distribution}
\label{App:DistChord}

\begin{figure}[hbt]
\begin{center}
\includegraphics[scale=0.5]{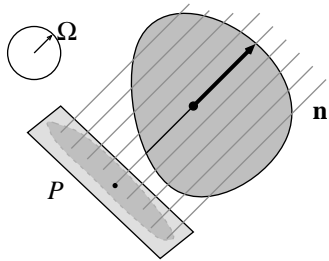}
\end{center}
\caption{Lines and chords}\label{Fig:lines}
\end{figure}

Let us change order of integration in \Eq{IntVSR}
\begin{equation}
\FI{\CB}(\Phi) = \frac{1}{V^2}\int_\Sph d\Omega \int_\CB d\ve{r} 
 \int_0^{\Rd(\ve{r},\ve{\Omega})}\!\! R^2 \Phi(R) dR  
\label{IntSVR}
\end{equation}
and for any unit vector $\ve{\Omega} \in \Sph$ to decompose spatial
integral $\int d\ve{r} = \iint dP\, dn$ on two integrals: along the axis 
$\ve{n} = \ve{n}_\Omega$ parallel to $\ve{\Omega}$ and on the plane 
$P=P_\Omega$ perpendicular to $\ve{\Omega}$ \Fig{lines}. 

After such decomposition \Eq{IntSVR} may be rewritten as
\begin{equation}
\FI{\CB}(\Phi) = \frac{1}{V^2}\int_\Sph d\Omega \int_{P_\Omega(\CB)} dP
 \int_0^{\Lh(p_\Omega,\ve{\Omega})}\!\!\! dn \int_0^{n} R^2 \Phi(R) dR,  
\label{IntSPRR}
\end{equation}
where $P_\Omega(\CB)$ is projection of $\CB$ on plane $P_\Omega$ and
$\Lh(p_\Omega,\ve{\Omega})$ is length of chord of line defined by direction 
$\ve{\Omega} \in \Sph$ and point $p_\Omega \in P_\Omega$, see \Fig{lines}.  

It is possible to define four-dimensional space $\T$ of (directed) 
lines \cite{Ken,San,Mat,Amb,Hit,Helg}, there each line is defined by point 
on sphere $\ve{\Omega} \in \Sph$ and point of intersection $p_\Omega$ with 
orthogonal plane $P_\Omega \perp \ve{\Omega}$. Here each line is represented 
twice with two opposite directions $\pm\ve{\Omega}$. It is possible to introduce 
space of undirected lines as a quotient space $\breve\T = \T/\{+1,-1\}$, but
in most expressions below for convenience of calculations is used $\T$.
The first two integrals in \Eq{IntSPRR} correspond to integration on this space
\begin{equation}
\FI{\CB}(\Phi) = \frac{1}{V^2}\int_{\T(\CB)}\!\! d\T 
 \int_0^{\Lh(l_\T)}\!\! dL \int_0^{L} R^2 \Phi(R) dR,  
\label{IntTRR}
\end{equation}
where $\T(\CB)$ is set of lines intersecting $\CB$, $\Lh(l_\T)$ is chord
length for a line $l_\T \in \T$, and $d\T$ is canonical (uniform and isotropic) 
measure on $\T$ unique defined up to constant multiplier by invariance 
with respect to translations and rotations \cite{Ken,San,Mat,Helg}.

Let us introduce linear functional
\begin{equation}
 \FIII{\CB}(\Upsilon) = \frac{1}{V[\T(\CB)]}\int_{\T(\CB)}\!\!\Upsilon[\Lh(l_\T)] d\T,
 \quad V[\T(\CB)] = \int_{\T(\CB)}\!\! d\T, 
\label{IntT}
\end{equation}
where $V[\T(\CB)]$ is normalization, {\ie} 4D volume of $\T(\CB)$ with respect to $d\T$.
It may be written also due to definition of $\int_{\T(\CB)}\! d\T$ above  
\begin{equation}
 V[\T(\CB)] = \int_\Sph d\Omega \int_{P_\Omega(\CB)}\!\! dP = 
 4 \pi \av{S_{P_\Omega(\CB)}}
\end{equation}
where $\av{S_{P_\Omega(\CB)}}$ is average surface of projection of body $\CB$. 
For convex bodies due to a Cauchy formula \cite{Ken,San} 
\begin{equation}
\av{S_{P_\Omega(\CB)}} = \frac{1}{4}S,
\label{Cauchy}
\end{equation}
where $S$ is surface area of $\CB$ and so
\begin{equation}
 V[\T(\CB)] = \pi S.
\label{NormFIII}
\end{equation}

\begin{Lemm}\label{LemTcld}
The linear functional $\FIII{\CB}(\Upsilon)$ defined by \Eq{IntT} may be rewritten
\begin{equation}
\FIII{\CB}(\Upsilon) = \int_0^\infty \Upsilon(x) \cld(x) dx 
\label{IntT2cld}
\end{equation}
where $\cld(x)$ is density of chord length distribution in body $\CB$ introduced
in \defn{defcld}.
\end{Lemm}
\begin{proof}
Similar with \lemm{LemV2dst} and \lemm{LemVSseg} the \Eq{IntT2cld} may be represented 
in rather trivial form $\FIII{\CB} = \cld \equiv T_\cld$ for generalized functions.
Here again the Heavyside step function $\Theta_L^\neg(x)$ selects 
subset in $\T(\CB)$ associated with chords shorter than given $L$.
The measure of integration $d\T$ in $\FIII{\CB}$ corresponds to uniform and
isotropic distribution of lines $l_\T$ in agreement with \defn{defcld} and so 
$\FIII{\CB}(\Theta_L^\neg) = \Prob(\Lh(l_\T) < L) 
= \int_0^L \cld(x) dx \equiv \Cld(l)$ and meets 
definition of density $\cld(x)$ for chord length distribution 
function $\Cld(x)$.
\end{proof}

The space of lines has nontrivial structure and an essential moment in proof
of \lemm{LemTcld} is correspondence of measure of integration $d\T$ in \Eq{IntT}
and measure describing distribution of lines in \defn{defcld}. 
This isotropic uniform distribution sometimes associated with concept of 
$\mu$-randomness \cite{Gil00,Han03,Kel71,Maz4J}. 

An alternative distribution of lines is often defined by uniform distribution 
of points inside the body $\CB$ and isotropic distribution of directions.
It is sometimes called $\nu$-randomness \cite{Gil00,Maz4J} or interior radiator
randomness \cite{Kel71}. It is different from $\mu$-randomness, because
the same line may be represented by any point on this line and direction
along this line. So measure of $\nu$-chord for given line in comparison with 
uniform case has extra multiplier proportional to length of the chord 
$\nu(l) \propto l\cld(l)$.
The precise expression for convex body is \cite{Gil00,Kel71,Maz4J}
\begin{equation}
  \nu(l) = \frac{l}{\av{l}}\cld(l) = \left(\frac{4V}{S}\right)^{-1}l\,\cld(l).
\label{mu2nu}
\end{equation}
The concept of $\nu$-randomness for chords is close with \defn{defseg}
of radii distribution and it is useful for some applications \cite{Kel71,Maz4J}.

Yet another distribution is produced by definition of line by pair of
points with independent uniform distributions inside $\CB$. It is sometimes 
called $\lambda$-randomness \cite{Gil00,Han03,Maz4J}. The measure for $\lambda$-chord
for three-dimensional case $\lambda(l) \propto l^4\cld(l)$. The normalizing
multiplier for $l^4$ may be directly calculated \cite{Gil00,MRD03,Mat,Dir,Maz4J}
\begin{equation}
\av{l^4} = \frac{12V^2}{\pi S}
\label{l4cld}
\end{equation}
and precise expression for convex body is \cite{Gil00,Maz4J}
\begin{equation}
  \lambda(l) = \frac{l^4}{\av{l^4}}\cld(l) = 
  \left(\frac{12V^2}{\pi S}\right)^{-1}l^4\cld(l).
\label{mu2lam}
\end{equation}
The concept of $\lambda$-randomness for chords has certain relation
with \defn{defdst} of distances distribution. 

Such different kinds of randomness illustrates necessity of rather pedantic
work with distribution of lines due to analogues of Bertrand paradox \cite{Ken,Prdx}
already mentioned earlier.

\begin{Lemm}\label{LemVS2T}
The functional \Eq{IntT} may be formally expressed via \Eq{IntVS}
\begin{equation}
 \FIII{\CB}\left(\int_0^x \Psi(l) dl\right) = \frac{4 V}{S} \FII{\CB}(\Psi).
\label{IntVS2T}
\end{equation}
\end{Lemm}
\begin{proof}
The derivation of \Eq{IntVS2T} is similar with transition from
\Eq{IntVSR} to \Eq{IntTRR} via \Eq{IntSVR} and \Eq{IntSPRR}
\begin{eqnarray*}
\frac{4 V}{S} \FII{\CB}(\Psi) &=& \frac{4 V}{S}
 \frac{1}{4\pi V}\int_\CB\int_\Sph \Psi[\Rd(\ve{r},\ve{\Omega})] d\ve{r}  d\Omega  
 = \frac{1}{\pi S} \int_\Sph\! d\Omega \int_\CB\! d\ve{r}\, \Psi[\Rd(\ve{r},\ve{\Omega})] 
 \\ &=& \frac{1}{\pi S} \int_\Sph\! d\Omega \int_{P_\Omega(\CB)} dP
 \int_0^{\Lh(p_\Omega,\ve{\Omega})}\!\!\! dn \int_0^n \Psi(l) dl = 
 \FIII{\CB}\left(\int_0^x \Psi(l) dl\right).
\end{eqnarray*}
\end{proof}

An application \Eq{IntT2cld} and \Eq{IntVS2seg} to \Eq{IntVS2T} produces
\begin{equation}
 \int_0^\infty\!\! \left(\int_0^x \Psi(l) dl \right) \cld(x) dx = 
  \frac{4 V}{S} \int_0^\infty \Psi(x) \seg(x) dx. 
\label{Icld2seg}
\end{equation}
An integration by parts of \Eq{Icld2seg} produces
\begin{equation}
 \int_0^\infty\!\! \left(\int_0^x \Psi(l) dl \right) \cld(x) dx = 
  -\frac{4 V}{S} \int_0^\infty\!\! \left(\int_0^x \Psi(l) dl \right) \seg'(x) dx. 
\label{Icld2seg2}
\end{equation}
The first equality in \Eq{der12} follows from \Eq{Icld2seg2}
\begin{equation}
 \seg'(x) = -\left(\frac{4 V}{S}\right)^{-1}\!\! \cld(x) = -\frac{1}{\av{l}}\cld(x).
\label{der2}
\end{equation}
It may be considered as generalized derivative due to \Eq{Icld2seg}. 
 
In \Eq{der2} was used Cauchy relation \Eq{Cau} $4V/S = \av{l}$. It is possible also
to find normalizing multiplier $n_\cld$ for $n_\cld \cld(x) = -\seg'(x)$ simply
using integration by parts
\begin{equation}
n_\cld^{-1} =  n_\cld^{-1}\int_0^\infty\!\!\seg(l) dl = 
- n_\cld^{-1}\int_0^\infty\!\!\seg'(l)l\,dl =  
 \int_0^\infty\!\! l \cld(l) dl =  \av{l}.
\label{avlcld}
\end{equation}
This derivation 
uses only the fact of
proportionality of $\cld(x)$ to derivative of another density function
and in further applications
for nonconvex case normalization with $\av{l}$ may be preferable due to
nontrivial proof of expression with volume and surface area.

\medskip

\begin{Lemm}\label{Lem2V2T}
The functional $\FI{\CB}$ \Eq{Int2V} may be expressed via $\FIII{\CB}$ \Eq{IntT}
\begin{equation}
  \FI{\CB}(\Phi) = \frac{S}{4V}\FIII{\CB}\Biggl(\int_0^x\!\!\biggl(%
\int_0^l\!  \frac{4 \pi}{V} r^2 \Phi(r)dr\biggl) dl \Biggl).
\label{Int2V2T}
\end{equation}
\end{Lemm}
\begin{proof}
The expression \Eq{Int2V2T} is straightforward combination of \Eq{IntVSR2V2}
and \Eq{IntVS2T}.
\end{proof}

Application of \Eq{IntT2cld} to \Eq{Int2V2T} produces
\begin{equation}
 \FI{\CB}(\Phi) = \frac{S}{4V}\int_0^\infty\!\!\left(\int_0^x\!\!%
 \int_0^l\!\frac{4 \pi}{V} r^2 \Phi(r)dr\,dl\right)\cld(x) dx.
\label{Int2V2cld}
\end{equation}

The \Eq{Int2V} and \Eq{Int2V2cld} produce for 
$\displaystyle\Phi(r) = V^2\frac{\varphi(r)}{4 \pi r^2}$
\begin{equation}
\int_\CB \int_\CB%
\frac{\varphi\bigl(|\ve{r}'-\ve{r}|\bigr)}{4 \pi |\ve{r}'-\ve{r}|^2} d\ve{r}\, d\ve{r}'
 = \frac{S}{4}\int_0^\infty\!\!\cld(x)%
\left(\int_0^x\!\!\!\int_0^p\!\!\varphi(r)dr\,dp\right) dx,
\label{VIntDir}
\end{equation}
in agreement with \Eq{IntDir} in \Sec{DirCh}.
It is also convenient sometimes to use expression
\begin{equation}
 \int_0^x\!\!\!\int_0^p\!\!\varphi(r)dr\,dp = 
 \int_0^x\!\!(x-r)\varphi(r)dr.
\label{I2I1}
\end{equation}

\section{Some equations for nonuniform case}
\label{App:nonun}

Let us consider a body with nonuniform density $\rho(\ve{r})$. 
Any such body $\NB$ may be treated as a convex one without lost of generality by
consideration of the convex hull $\CB$ and assignment of zero density to 
the complement $\CB\setminus\NB$.

Here is considered two cases corresponding to choice of $\Delta_{\ve{r},\ve{r}'}$
in \Eq{IntDirNonUn}.

\subsection{Distance between points $\Delta_{r,r'}=|r'-r|$}
\label{App:nonun_dist}

A direct analogue of \Eq{Int2Vcor} is 
\begin{eqnarray}
 \iint \rho(\ve{r}) \rho(\ve{r}')%
 \frac{\varphi\bigl(|\ve{r}'-\ve{r}|\bigr)}{4\pi|\ve{r}'-\ve{r}|^2} d\ve{r}\, d\ve{r}' 
 &=& \int_0^\infty \cor(x) \varphi(x)  dx \nonumber \\
 &=& \int_0^\infty\!\!\cor''(x)%
 \left(\int_0^x\!\!\!\int_0^p\!\!\varphi(r)dr dp\right) dx,
\label{Int2Mcor}
\end{eqnarray}
there second equality produced by two integrations by parts. So up to normalization
with $\int_0^\infty \cor''(x) dx = \cor'(0)$ 
it is possible to use a formal (``generalized'' \cite{Str01,Han03}) chord length 
distribution $\acute\cld(l) = \cor''(l) /\cor'(0)$ for calculation of integrals
like \Eq{Int2Mcor}.

Finally
\begin{equation}
 \iint \rho(\ve{r}) \rho(\ve{r}')%
 \frac{\varphi\bigl(|\ve{r}'-\ve{r}|\bigr)}{4\pi|\ve{r}'-\ve{r}|^2} d\ve{r}\, d\ve{r}' =
 \acute{C}_\cld \int_0^\infty\!\!\acute\cld(x)%
 \left(\int_0^x\!\!\!\int_0^p\!\!\varphi(r)dr dp\right) dx,\quad \acute{C}_\cld=\cor'(0).
\label{Int2Mfcld}
\end{equation}

There is yet another way to express $\acute{C}_\cld$. If to consider 
$\varphi(l)=4 \pi l^2$, then from \Eq{Int2Mfcld} follows
$$
M^2 = \iint \rho(\ve{r}) \rho(\ve{r}') d\ve{r}\, d\ve{r}' = 
 \acute{C}_\cld \int_0^\infty\!\!\acute\cld(x)%
 \left(\int_0^x\!\!\!\int_0^p\!\! 4 \pi r^2 dr dp\right) dx =
 \frac{\pi}{3}\acute{C}_\cld \int_0^\infty\!\!x^4\acute\cld(x)dx.
$$
where $M \equiv \int \rho(\ve{r}) d\ve{r}$ is mass of the body. Let us also
denote $\acute{\av{l^4}} \equiv \int_0^\infty\!\!x^4\acute\cld(x)dx$, then
\begin{equation}
 \acute{C}_\cld = \frac{3 M^2}{\pi \acute{\av{l^4}}}.
\label{acCM}
\end{equation}

For case of constant unit density $M=V$ and due to \Eq{l4cld} $\av{l^4} = 12V^2/(\pi S)$. 
So $C_\cld = S/4$ in agreement with \Eq{VIntDir} and Cauchy formula \Eq{Cauchy}. 

\subsection{``Optical'' length $\Delta_{r,r'}=\opt{r}{r'}$}
\label{App:nonun_opt}

In some applications instead of distance between points $R=|\ve{r}'-\ve{r}|$ in 
$\varphi(R)$ it is necessary to use an ``optical width'', {\ie}
integral on density along line between points
\begin{equation}
 \opt{\ve{r}}{\ve{r}'} = \smallint_{\ve{r}}^{\ve{r}'}\!\rho\,d\ell =
 |\ve{r}'-\ve{r}|\int_0^1\rho\bigl(\ve{r}+(\ve{r}'-\ve{r})x\bigr)dx
\label{OptLen}
\end{equation}
and to consider functional
\begin{equation}
 \FJ(\varphi) =
 \iint \rho(\ve{r}) \rho(\ve{r}')%
 \frac{\varphi(\opt{\ve{r}}{\ve{r}'})}{4\pi|\ve{r}'-\ve{r}|^2} d\ve{r}\, d\ve{r}'
\label{FJ}
\end{equation}
Let us introduce new variable and rewrite integral on $d\ve{R}$ 
using spherical coordinates like in \Eq{IntVSR}
\begin{equation}
 \FJ(\varphi) = \frac{1}{4\pi}%
 \int_{\R^3}\!\!d\ve{r}\int_\Sph d\Omega\,\rho(\ve{r})\!\int_0^{\Rd(\ve{r},\ve{\Omega})}%
 \!\!\rho(\ve{r}+x\ve{\Omega})\varphi(\opt{\ve{r}}{\ve{r}+x\ve{\Omega}})dx.
\label{IntMSR} 
\end{equation}
where $\Rd(\ve{r},\ve{\Omega})$ is lenght of (maximal) radius, 
$\opt{\ve{r}}{\ve{r}+x\ve{\Omega}} = \int_0^x\rho(\ve{r}+l\ve{\Omega})dl 
 \equiv s(x)$, $s'(x)=\rho(\ve{r}+x\ve{\Omega})$. Last
integral has form $\int_0^l s'(x) \rho(s(x)) dx$ and may be rewritten as
$\int_0^{s(l)} \rho(s) ds$.
\begin{equation}
 \FJ(\varphi) = \frac{1}{4\pi} \int_{\R^3}\!\!d\ve{r}\int_\Sph d\Omega%
\,\rho(\ve{r})\!\int_0^{w(\ve{r},\ve{\Omega})}\!\!\varphi(w)dw, 
\label{IntMSW}
\end{equation}
where $w(\ve{r},\ve{\Omega})=\opt{\ve{r}}{\ve{r}+\Rd(\ve{r},\ve{\Omega})\ve{\Omega}}=
\int_0^{\Rd(\ve{r},\ve{\Omega})}\rho(\ve{r}+l\ve{\Omega})dl$
is ``optical'' radius length.

It is possible to use analogue of integration in \Eq{IntSPRR}
with axis $\ve{n} \parallel \ve{\Omega}$ and plane $P \perp \ve{\Omega}$
\begin{equation}
\FJ(\varphi) = \frac{1}{4\pi}\int_\Sph d\Omega\!\!\int_P dP\!\!\int dn%
\,\rho(\ve{r})\!\int_0^{w(\ve{r},\ve{\Omega})}\!\!\varphi(w)dw,
\quad \ve{r} = \ve{r}_P+n\ve{\Omega},
\label{IntSPNW}
\end{equation}
where point $\ve{r}_P$ is intersection (projection) of line representing path 
of integration with plane $P$ (see \Fig{lines}) and range of $n$ corresponds to 
variation of $\ve{r}$ along whole chord in convex span of a body. 

If to consider last integral in \Eq{IntSPNW} as some function $F(w(n))$,
two last integrals have form $\int w'(n) F(w(n)) dn = \int F(w) dw$ and so
we have analogue of \Eq{IntTRR}
\begin{equation}
\FJ(\varphi) = \frac{1}{4\pi}\int_\T\!\!d\T\int_0^{\Wd(l_\T)}dw\int_0^x \varphi(x) dx,
\label{IntTWW}
\end{equation}
where $\Wd(l_\T)=\opt{\ve{r}_{\min}}{\ve{r}_{\max}}(l_\T)$ is ``optical'' chord length. 

\begin{Lemm}\label{Lem2V2ocld}
Let us now introduce ``optical'' chord length distribution $\tilde\cld(x)$, {\ie}
to any chord of line intersecting (convex hull of) a body in two points $\ve{r}_1$ 
and $\ve{r}_2$ instead of $|\ve{r}_2-\ve{r}_1|$ is assigned ``optical'' length 
$\opt{\ve{r}_1}{\ve{r}_2}$. The integral \Eq{FJ} may be expressed
\begin{equation}
 \FJ(\varphi) = \tilde{C}_\cld\int_0^\infty\!\!\tilde\cld(x)%
 \left(\int_0^x\!\!\!\int_0^p\!\!\varphi(r)dr\,dp\right) dx.
\label{Int2V2ocld}
\end{equation}
\end{Lemm}
\begin{proof}
The \Eq{Int2V2ocld} follows from \Eq{IntTWW} and
it is complete analogue of derivation \Eq{Int2V2cld} from \Eq{IntTRR}
in \App{CalcDist}. In both cases is used the same integral of some function 
$\Fd(l_\T)$ over space of lines $\T$ with the same measure $d\T$ and particular
method of calculation of $\Fd$ does not matter. The $\cld(l)$ in \Eq{Int2V2cld}
is (joint) density for $\Fd(l_\T)=\Lh(l_\T)$ and the $\tilde\cld(l)$ in \Eq{Int2V2ocld} 
is (joint) density for $\Fd(l_\T)=\Wd(l_\T)$. 

There is unessential difficulty with constant multiplier $\tilde{C}_\cld$, due 
to lack of simple analogue of Cauchy formula \Eq{Cauchy} for average surface
used in normalization of the integral on $\T$ in \Eq{NormFIII}. An alternative
way of calculation of $\tilde{C}_\cld$ is represented below.
\end{proof}

Let us introduce quantity
\begin{equation}
 \G \equiv 
 \iint \frac{\rho(\ve{r})\rho(\ve{r}')}{4\pi|\ve{r}'-\ve{r}|^2} d\ve{r}\, d\ve{r}' =
 \int_0^\infty \cor(x) dx
\label{defG}
\end{equation}
then for \Eq{Int2V2ocld} with $\varphi(l)=1$
$$
 \G = \tilde{C}_\cld \int_0^\infty\!\!\tilde\cld(x)%
 \left(\int_0^x\!\!\!\int_0^p\!\! dr dp\right) dx =
 \frac{1}{2}\tilde{C}_\cld \int_0^\infty\!\!x^2\tilde\cld(x)dx =
 \frac{1}{2}\tilde{C}_\cld \tilde{\av{l^2}}
$$
and
\begin{equation}
 \tilde{C}_\cld = \frac{2G}{\tilde{\av{l^2}}}
\label{tilC}
\end{equation}

Finally
\begin{equation}
 \iint \rho(\ve{r}) \rho(\ve{r}')%
 \frac{\varphi(\opt{\ve{r}}{\ve{r}'})}{4\pi|\ve{r}'-\ve{r}|^2} d\ve{r}\, d\ve{r}' =
 \tilde{C}_\cld \int_0^\infty\!\!\tilde\cld(x)%
 \left(\int_0^x\!\!\!\int_0^p\!\!\varphi(r)dr dp\right) dx,
\quad \tilde{C}_\cld = \frac{2G}{\tilde{\av{l^2}}}.
\label{Int2Mocld}
\end{equation}

\section{Average path length} 
\label{App:pathlen}

Here is presented simple geometrical proof of equality of average path
length inside a body and average chord length in uniform isotropic
case.

\begin{figure}[hbt]
\begin{center}
\includegraphics[scale=0.5]{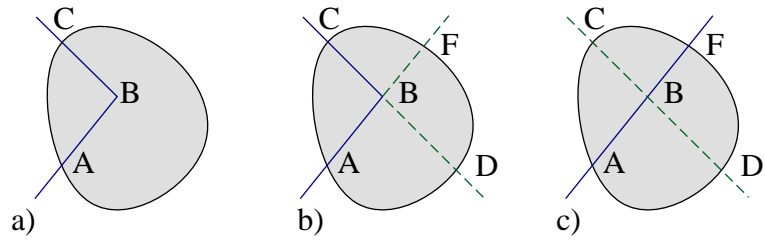}
\end{center}
\caption{Illustration for average path length}\label{Fig:pathlen}
\end{figure}

On the \Fig{pathlen}a is depicted some path $ABC$ with one kink inside
a body. Let us also consider path $FBD$ produced from $ABC$ by central
symmetry with respect to point $B$, \Fig{pathlen}b. The sum of lengths
of two paths $FBD$ and $ABC$ is $|AB|+|BC|+|FB|+|BD|$ and 
coincides with sum for two chords $AF$ and $CD$ \Fig{pathlen}c.

This method let us get rid of one kink and after few such steps to consider 
average length of straight lines instead of paths. The necessary
condition for such a proof --- is isotropic and uniform distribution of 
paths to ensure equal probability (density) of $ABC$ and $FBD$.

\end{document}